\documentclass[aps,pre,twocolumn,showpacs,superscriptaddress]{revtex4-1}
\usepackage{mathrsfs}
\usepackage{graphicx}
\usepackage{bm}
\usepackage{amsmath,amssymb}

\usepackage{color}

\newcommand{\av}[1]{\left\langle #1 \right\rangle}

\begin{document}
\title{Anisotropic thermophoresis}
\author{Zihan Tan}\email{z.tan@fz-juelich.de}
\affiliation{Theoretical Soft-Matter and Biophysics, Institute of
Complex Systems, Forschungszentrum J\"ulich, 52425 J\"ulich, Germany}
 \author{Mingcheng Yang}\email{mcyang@iphy.ac.cn}
 \affiliation{Beijing National Laboratory for Condensed
 Matter Physics and Key Laboratory of Soft Matter Physics, Institute
 of Physics, Chinese Academy of Sciences, Beijing 100190, China}
 \affiliation{University of Chinese Academy of Sciences, Beijing 100049, China}
\author{Marisol Ripoll}\email{m.ripoll@fz-juelich.de}
\affiliation{Theoretical Soft-Matter and Biophysics, Institute of
Complex Systems, Forschungszentrum J\"ulich, 52425 J\"ulich, Germany}
\date{\today}

\begin{abstract}
  Colloidal migration in temperature gradient is referred to as
  thermophoresis. In contrast to particles with spherical shape, we
  show that elongated colloids may have a thermophoretic response that
  varies with the colloid orientation. Remarkably, this can translate
  into a non-vanishing thermophoretic force in the direction
  perpendicular to the temperature gradient. Oppositely to the
  friction force, the thermophoretic force of a rod oriented with the
  temperature gradient can be larger or smaller than when oriented
  perpendicular to it. The precise anisotropic thermophoretic behavior
  clearly depends on the colloidal rod aspect ratio, and also on its
  surface details, which provides an interesting tunability to the
  devices constructed based on this principle. By means of mesoscale
  hydrodynamic simulations, we characterize this effect for different
  types of rod-like colloids.
\end{abstract}


\maketitle

\section{Introduction}
The migration of particles in the presence of a temperature gradient
was first described in the XIX century~\cite{ludwig56,soret79} and it
is generally known as thermal diffusion or Soret effect.  The study of
the Soret effect in liquid mixtures is usually referred to as {\em
  thermodiffusion}~\cite{wieg04}. In these mixtures, all
relevant particles sizes are similar and within the nanometer scale.
The same physical principle applied to colloids is typically referred
to as {\em thermophoresis}~\cite{piazza08r,wuerger10r}. Colloids can
have sizes up to the micrometer scale and the surrounding solvent
particles can be orders of magnitude smaller.  Several factors are
known to importantly influence the thermodiffusive behavior of these
systems like the colloid size, mass, average temperature,
concentration~\cite{wieg04,piazza08r}.  In fluid mixtures the effect
of the particle moment of inertia has been extensively
studied~\cite{plathe00,koeh01,gal03}, concluding that increased moment
of inertia facilitates migration. Rodlike colloids have been
experimentally investigated~\cite{blanco11,wang13} and characterized
as a function of their electrostatic interactions.  However, no
systematic study has been yet done to investigate the thermophoretic
properties of colloids as a function of their shape.

Most colloids show thermophobic behavior (average migration to the
cold areas), although thermophilic colloids have also been frequently
reported~\cite{braun06,piazza06}. In both cases, the colloid migration
is characterized by a unique value of the so-called {\em thermal
  diffusion factor} which, by convection, is positive for thermophobic
colloids and negative for thermophilic. The sign and the value of the
thermodiffusion factor is determined by the colloid size and
properties of the colloid solvent interactions, and it can be modified
or even reversed for example by changing average temperature,
concentration, pH or solvent composition.  
But the question that arises is how is the thermophoretic behavior of
elongated particles. 
A valuable and interesting parallelism can be made here between
thermophoretic and friction forces.  An elongated object moving in a
fluid along its axis is known to experience frictional force
$\gamma_\|$, which is typically much smaller than the friction
experienced by the same rod moving perpendicular to its axis
$\gamma_\perp$. This is a well-know fact, which in the case of a
shish-kebab model of adjacent beads has been calculated to be
$\gamma_\perp = 2 \gamma_\|$ for aspect ratios larger than
20~\cite{doi}.  
It is therefore intuitive, that an elongated object with its axis
aligned with a temperature gradient, will not have the same
thermophoretic response as when the axis is perpendicular to the
gradient. 
Hence and in contrast to colloids with spherical symmetry, colloids
with an anisotropic shape should be characterized by two or more
thermal diffusion factors, which is the main concept of the {\em
  anisotropic thermophoresis}~\cite{yang14turb}.

Practical applications of thermodiffusion have developed over several
decades and are currently in a significant expansion stage. Some
relevant examples are crude oil characterization~\cite{ghorayeb03},
separation techniques~\cite{gidd93}, strong components
accumulation in prebiotic conditions~\cite{braun07pnas,niether16}, the
precise characterization of proteins, for which thermophoresis can even
distinguish betweeen different binding states~\cite{jerabek14}; also 
various application in microfluidics~\cite{piazza10,16tratchet}, or the 
fabrication of synthetic microswimmers~\cite{jiang10,yang11dimer,yang14janus}. 
The traditional versatility of thermophoresis is therefore importantly
increased by considering different shaped objects.

In this work we investigate the anisotropic thermophoretic properties
of colloidal rods by means of hydrodynamic computer simulations. We
study the dependence of the thermophoretic forces for moving rods and
for fixed rods at different orientations, with various aspect
ratios and surface properties.  The anisotropic effect can for example
be reversed by changing the surface rugosity, which can be understood
in terms of the associated temperature properties of the fluid in the
vicinity of the colloid.  Interestingly, this anisotropy can induce a
thermophoretic effect non-aligned with the temperature gradient. This
contribution to the thermophoretic force perpendicular to the
temperature gradient has already shown to be the basic mechanism that
allows the construction of thermophoretic turbines, which move in the
presence of an external temperature gradient~\cite{yang14turb}. Other
promising applications are not yet explored, but are certainly to be
expected, and are specially promising in microfluidic devices where
significantly large and well-localized temperature gradients can
be generated and precisely controlled in time and space. Applications
of anisotropic thermophoresis in the presence of external temperature
gradients offer then the possibility of engineering devices able to
harvest waste heat energy.

\section{Simulation method} 
\label{sec:mpc}
In order to be able to bridge the time and length scales of the
colloidal rods and the surrounding solvent particles, a hybrid
description of the suspension is employed.  The solvent is simulated
by an efficient technique known as multiparticle-collision dynamics
(MPC)~\cite{kap99,kap00,kapral_review}, while the colloidal rod is
described with a Molecular Dynamics simulation model.  In MPC, the
solvent is represented by $N$ point particles whose dynamics takes
place in two sequential steps. One is the streaming step, at which all
point particles move ballistically for a certain collision time
$h$. In the collision step, particles are grouped into cubic collision
cells of size $a$, where particles interchange momentum by performing a
rotation by a certain angle $\alpha$ of the particle velocity relative
to the corresponding cell center of mass velocity.  MPC accounts for
thermal fluctuations, and conserves energy and linear momentum both
globally and locally~\cite{ihl01,pre05,tuz06} by construction.
Angular momentum is not conserved for this MPC
implementation~\cite{goe07,nog07}.  However, recent work has provided
evidence that angular momentum conservation in MPC fluid does not
influence the hydrodynamics of phoretic flows~\cite{yang15amc}. 
Simulations are performed with an average of $\rho=10$ particles per
collision cell, the collision time step $h=0.1$, the rotation angle
$\alpha=130^\circ$, and the average temperature of the solvent
$k_{B}\overline{T}=1$. These parameters yield the viscosity
$\eta=8.7$ according to the theoretical
predictions~\cite{kap00,ihl01,yeo03,yeo05a,ihl05b,nog07b,winkler09},
which is in close agreement with simulation results \cite{huang15}.
In the following, all quantities are expressed in terms of the MPC
units, where the units of length, mass and energy are separately
imposed as $a$, $m$ and $k_{B}\overline{T}$, such that the time unit
is $a\sqrt{m/k_{B}\overline{T}}$.  The temperature gradient is applied
in the $z$ direction with a boundary
thermostat~\cite{luese12jcp,yang13flow,plathe97}, ensuring that heat
conduction is correctly accounted for.

\begin{figure}[h]
\includegraphics[width=0.45\textwidth]{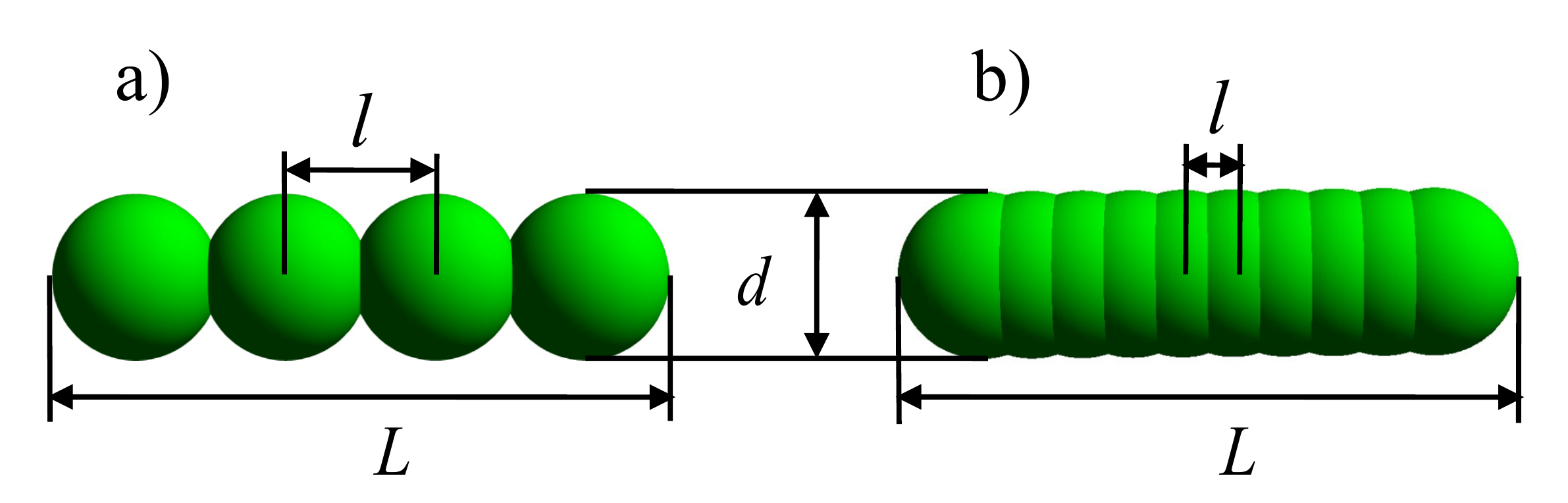}
  \caption{Sketch of the ``shish-kebab'' model of a colloidal rod of
    aspect ratio $L/d=3.7$ with variable number of beads $N$, and
    variable inter-bead separation $l$.  (a)~``\textit{Rough}'' rod
    with $N=4$ beads and $l=0.9d$. (b)~``\textit{Smooth}'' rod with
    $N=10$ beads and $l=0.3d$.}
  \label{model}
\end{figure}

The rod is constructed with the ``shish-kebab'' model built by $N$
connected beads in a linear disposition as shown in Fig.~\ref{model}.
The excluded volume interactions between colloid and solvent are
performed via MD with Lennard-Jones (LJ) type
potentials~\cite{vliegent99,luese12jpcm}
\begin{equation}\label{lj}
U_k(r) =
4\epsilon\left[\left(\frac{d}{2r}\right)^{2n}
  -\left(\frac{d}{2r}\right)^{n}+c\right], \  r\leq r_c. 
\end{equation}
Here $r$ is the distance between the bead center and the fluid
particle, $\epsilon$ refers to the potential intensity, $d$ to the
bead diameter, and $n$ is a positive integer describing the potential
stiffness. The repulsive or attractive LJ potentials are obtained by
considering $c=1/4$ or $c=0$ respectively, together with the adequate
cutoff distance $r_c$. The repulsive and attractive potentials with
stiffness $n=6$ will be denoted as {\em r6} and {\em a6} respectively,
and similar for other $n$ values. The bead diameter is taken as

$d=4a$, and $\epsilon=k_{B}\overline{T}$.  The mass of each
  bead is chosen such that the rod is neutrally buoyant, although
  results are not really depending on this value.  The equations of
motion of the beads and interacting fluid are integrated with a
velocity-Verlet MD algorithm. %
The rod length $L$ is given by $L=d+(N-1)l$, such that the aspect
ratio $L/d$ can be fixed for more than one combination of $N$ and $l$.
Besides the particular solvent-bead potential, the rod surface
properties are going to be determined by the interbead separtion,
which we characterize with $l/d$, the rugosity parameter. As can be
seen in Fig.~\ref{model}, a rough rod is obtained when $l\simeq d$,
while a smooth rod is obtained when $l\ll d$. In order to prevent the
penetration of the fluid inside the rod, the maximum value of the
rugosity parameter we employ is $l/d = 0.9 $.  We consider a
three-dimensional box with periodic boundary conditions changing from
$3$ to $4$ times the rod length $L$.  Two types of simulations
  are performed. Simulations with fixed rods only need the collod
  solvent interactions to be specified. Simulation with freely
  rotating rods and fixed central of mass, in which the rod motion is
  additionally accounted with rigid body dynamics~\cite{allen}.

\section{Results}
\begin{figure}[h!]
\includegraphics[width=0.45\textwidth]{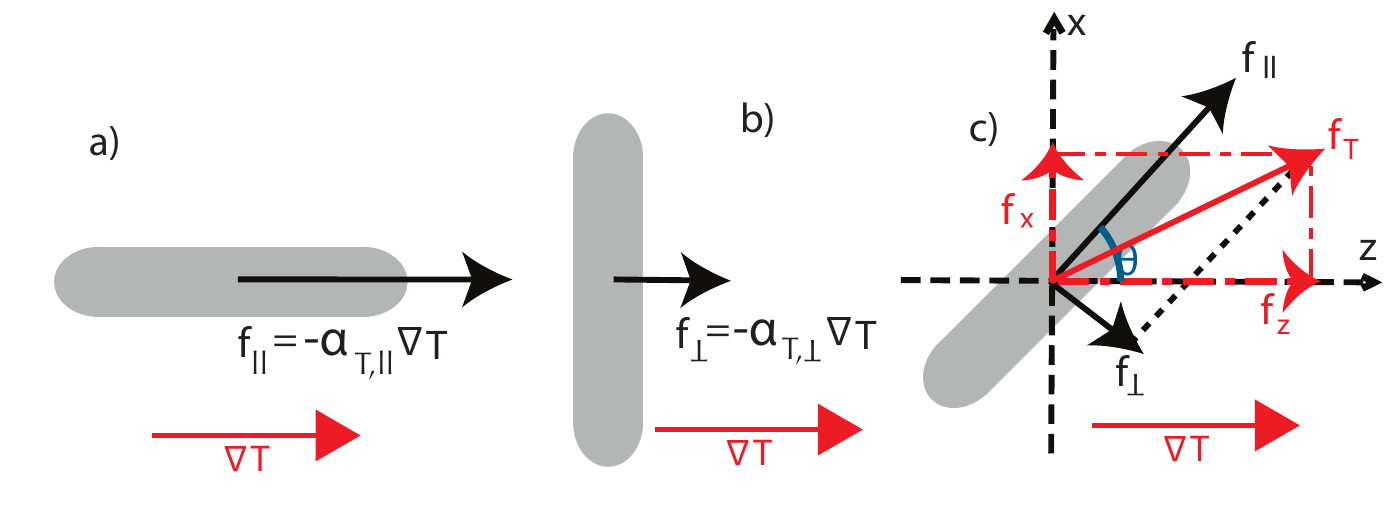}
\caption{Sketch of the thermophoretic force of a colloidal rod fixed
  with different orientations with respect to the temperature
  gradient: a) parallel, b) perpendicular, c) with an arbitrary
  $\theta$ angle. \label{fig:t-anisotrop}}
\end{figure}

The migration of a particle in a temperature gradient is driven by the
thermophoretic force ${\bf F}_T$~\cite{piazza05,yang12forces,yang12drift}. 
This force arises from the interactions between the particle and the
surrounding inhomogeneous fluid environment. In the case of
spherically symmetric colloids, this force is characterized by a
unique value of the so-called {\em thermodiffusion factor},
$\alpha_T$, as
\begin{equation}
\textbf{F}_T=-\alpha_Tk_B\nabla T,
\label{eq4}
\end{equation}
with $k_B$ the Boltzmann constant, and $\nabla T$ the temperature
gradient.  There is though no reason to expect that a unique factor
would still describe the thermophoretic behavior of an non-spherically
symmetric particle. 
A straightforward generalization can be written in terms of the {\em
  thermodiffusion tensor} $\Lambda_T$ as
\begin{equation}
  \textbf{F}_T=-\Lambda_T \cdot \nabla T.
  \label{eq8}
\end{equation}
In a very general case, colloids with arbitrary shape and arbitrary
surface properties can be defined by a symmetric tensor with
independent coefficients. Of particular interest is the case of an
homogeneous colloidal rod with cylindrical symmetry. In this case, two
independent coefficients are expected to be enough to determine the
thermodiffusion tensor as
\begin{equation}
  \Lambda_T=\alpha_{T,\|}\hat{\textbf{u}}\hat{\textbf{u}}+
\alpha_{T,\perp}(\hat{\textbf{I}}-\hat{\textbf{u}}\hat{\textbf{u}}),
  \label{eq7}
\end{equation}
with $\hat{\textbf{u}}$ the unit vector of long axis of rod. Here
$\alpha_{T,\|}$ is the thermodiffusion factor of the long rod axis, or
equivalently the thermophoretic factor that characterizes a rod with
the long axis aligned with the temperature gradient, as displayed in
Fig.~\ref{fig:t-anisotrop}a. Reciprocally, $\alpha_{T,\perp}$ is the
thermodiffusion factor of the short rod axis, or of a rod with the
long axis oriented perpendicular to the temperature gradient (see
Fig.~\ref{fig:t-anisotrop}b). 

\begin{figure}[h!]
\includegraphics[width=0.35\textwidth]{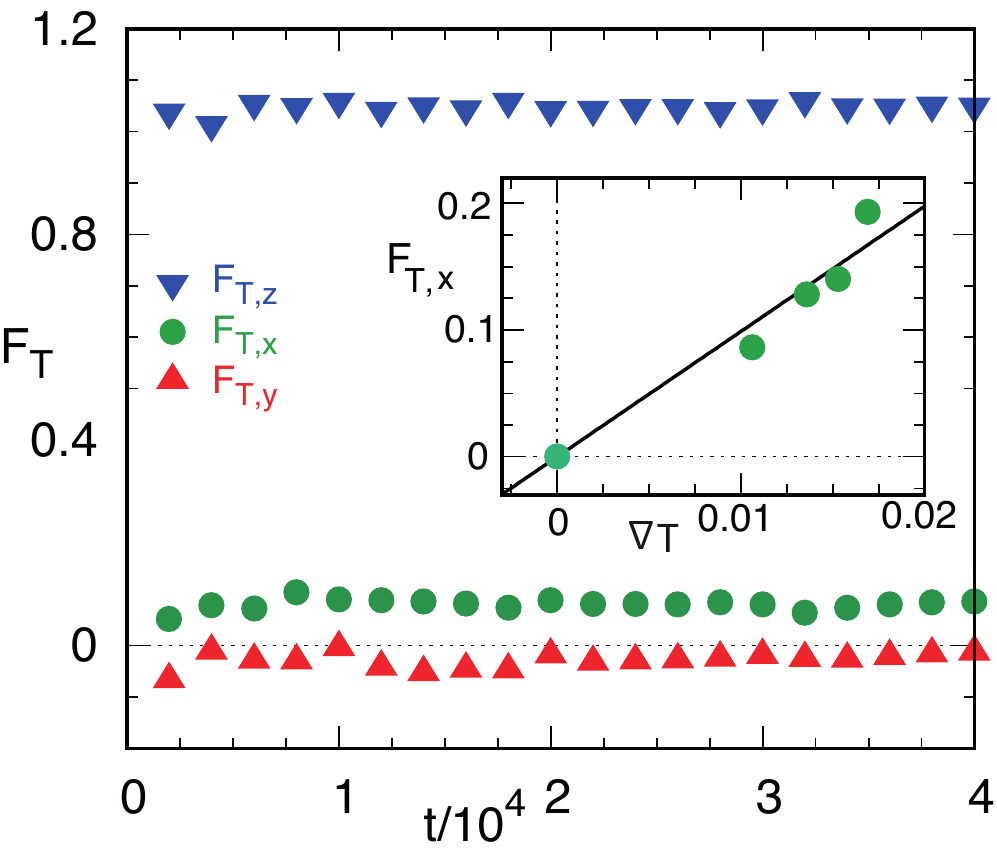}
\caption{Thermophoretic force obtained from simulations as a time
  average for a fixed smooth rod ($l/d= 0.3$), of aspect ratio 
  $L/d=3.7$, and interaction potential $r6$.  The rod is fixed with 
  an angle $\theta =45^\circ$ with respect to $\nabla T$. 
  The force perpendicular to $\nabla T$, $F_{T,x}$, is non-vanishing as 
  predicted by Eq.~(\ref{perp}). The inset shows the value of $F_{T,x}$ 
  for various values of $\nabla T$.
  \label{fig:ft45}}
\end{figure}

\paragraph{Linear decomposition of the thermophoretic force.-}
The thermophoretic force acting on a rod with arbitrary orientation
can be determined by the linear superposition of the two components
with orthogonal thermodiffusion factors ${\bf F}_T= {\bf F}_\parallel+{\bf
  F}_\|$. The temperature gradients along the long and short axis are
respectively $\nabla T_l=cos\theta\nabla T$ and $\nabla
T_s=sin\theta\nabla T$, with $\theta$ the angle between the particle
long axis and the temperature gradient, as displayed in
Fig.~\ref{fig:t-anisotrop}c. The total force can then be expressed in
terms of its components, parallel and perpendicular to the temperature
gradient ${\bf F}_T= {\bf F}_z+{\bf F}_x$ as,
\begin{eqnarray}
{\bf F}_{z}&=&-\left(\alpha_{T,\perp}\sin^{2}\theta+\alpha_{T,\|}\cos^{2}\theta\right)k_{B}\nabla
T,\label{para}\\
{\bf F}_{x}&=&\left(\alpha_{T,\perp}-\alpha_{T,\|}\right)\sin\theta\cos\theta
k_{B}|\nabla T|{\bf n}_{x}\label{perp} 
\end{eqnarray}
where ${\bf n}_{x}$ is the unit vector perpendicular to $\nabla T$.
Equation~(\ref{perp}) comes as a straightforward result of the
tensorial character of the thermodiffusion tensor in Eq.~(\ref{eq8}),
and strikingly implies that a non-vanishing thermophoretic force
exists in the direction perpendicular to the temperature
gradient. This force can in fact be easily measured in our simulations
as shown in Fig.~\ref{fig:ft45}, such that it could also be measured
experimentally~\cite{jiang09,helden15}.  The measured force
perpendicular to the temperature gradient is cleary non-vanishing, and
it increases linearly with the applied tempeature gradient, as
expected from Eq.~(\ref{perp}). This nicely confirms the tensorial
character of the thermophoretic effect for objects without spherical
symmetry.

\begin{figure}[h!]
\includegraphics[width=0.7\columnwidth]{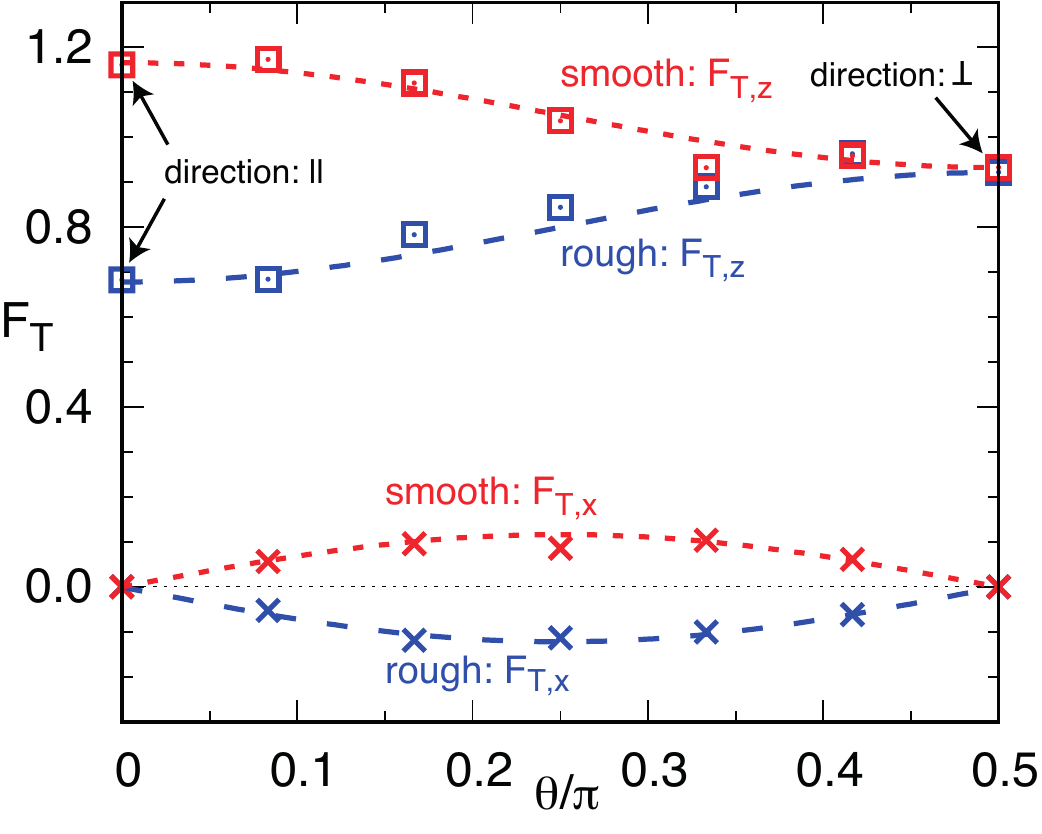}
\caption{Thermophoretic force for fixed smooth ($l/d= 0.3$) and 
  rough ($l/d= 0.9$) rods with  $L/d=3.7$ and {\em r6}.
   The angle $\theta$ denotes the rod orientation with respect to
  $\nabla T$, as sketched in Fig.~\ref{fig:t-anisotrop}c.  The force
  $F_{T,z}$ is measured in the direction paralle to $\nabla T$ and
  $F_{T,x}$ perpendicular to it. Symbols correspond to simulation
  results and lines to the expressions in Eq.~(\ref{para}) and
  (\ref{perp}).
  \label{fig:FT_angle}}
\end{figure}

Simulations with single rods fixed by an angle $\theta $ with respect
to the temperature gradient are performed for different orientations,
as shown in Fig.~\ref{fig:FT_angle}.  The values of $\alpha_{T,\|}$
and $\alpha_{T,\perp}$ can be obtained by fitting the expressions
Eqs.~(\ref{para}) and (\ref{perp}) to the simulation results, or more
efficiently, just by fixing the rod parallel ($\theta=0$) or
perpendicular($\theta=\pi/2$) to the temperature gradient.  The linear
decomposition of the thermophoretic force in Eqs.~(\ref{para}) and
(\ref{perp}) is clearly verified by these simulation results.

Results in Fig.~\ref{fig:FT_angle} also show that, depending on the
simulated rods, $F_{T,\|}$ can be larger or smaller than
$F_{T,\perp}$, such that the force perpendicular to the gradient,
$F_{T,x}$, can appear in both directions. This is in strong contrast to the
friction force, which is always larger for rods oriented perpendicular
to the flow than for those parallel to it.  Hence, the anisotropic
thermophoretic effect is fundamentally different from the anisotropy
of the translational friction.

\paragraph{Does thermophoretic anisotropy  induce orientation?}
As just discussed, elongated particles fixed aligned with the
temperature gradient or perpendicular to it, can experience
well-differentiated thermophoretic forces.  Nevertheless, a freely
moving rod in a temperature gradient, suffers a force exerted on one
half of the rod which is exactly the same as in the other half, such
that there is no net torque on the particle.  For the same reason,
friction forces are also known not to induce any orientation effects
on elongated particles, in spite of their anisotropy. Orientation is
induced only in the case that the flow field is in itself not
homogeneous, as it is the case of a shear flow.  
The thermophoretic anisotropy does therefore induce no particle
alignment. A freely rotating rod in a temperature gradient, will then
change its orientation only due to stochastic interactions. 
Simulations allowing particle rotation confirm this statement for the
two parameter sets in Fig.~\ref{fig:FT_angle}, and the measured
thermodiffusion factors verify that $\alpha_{T,free}
=\av{\alpha_{T}}_{\theta} \simeq \alpha_{T,\|}|_{\theta=45}$.  

While alignment refers to the first moment of the induced
  orientation, we could wonder if the effect exist in higher moment
  orders.  It is therefore interesting to investigate if the presence
of a external temperature gradient could modify the particle
rotational diffusion.  We measure the rotational diffusion coefficient
by characterizing the long time behaviour of the mean squared
orientation displacement.  For computational efficiency, these
 simulations consider rods composed of smaller beads, $d=2a$. The
normalizing factor $D_r^0\simeq 8 \times 10^{-4}$ is obtained for the
rough rods in the absence of temperature gradient. In the
  presence of non-vanishing temperature gradients, simulations are
performed by keeping the rod center of mass fixed in the middle of the
simulation box where the solvent average temperature is
$k_B\overline{T} = 1$.
\begin{figure}[h!]
\includegraphics[width=0.25\textwidth,angle=-90]{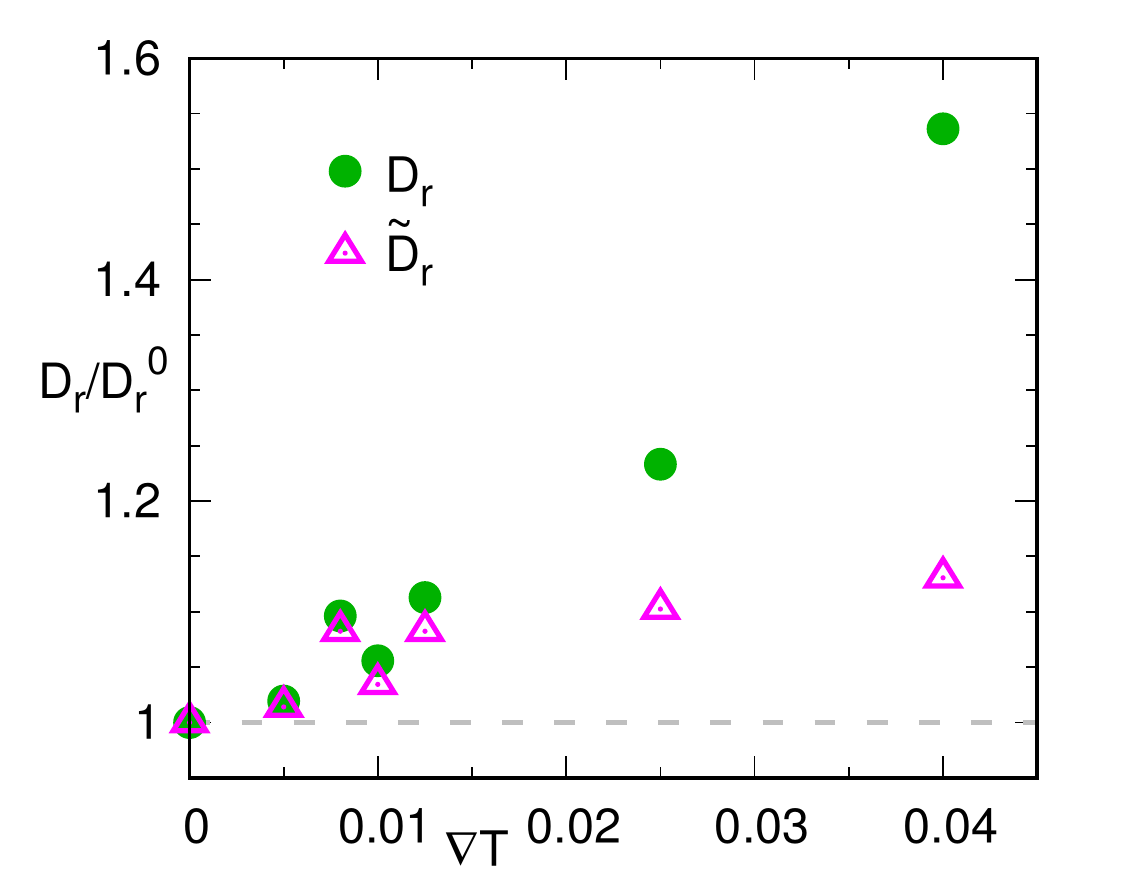}
\caption{Normalized diffusion coefficient for a freely rotating rod
  (with $l/d= 0.3$, $L/d=3.7$, and $r6$) at various temperature
  gradients.  Bullets correspond to the direct simulation
  measurements $D_r$, and triangles to the coefficient rescaled with the
  density correction $\tilde{D}_r$. \label{fig:Dr}}
\end{figure}
The simulation data in Fig.~\ref{fig:Dr} show to increase with the
applied temperature, but this is mainly due to the ideal gas character
of the employed MPC fluid. Given a spatial linear increase of the
temperature, the related position dependent density will be only
approximately linear~\cite{luese12jcp}. The density in the middle of
the simulation box will then differ from the average one, which will
affect the rotational diffusion coefficient.  If we consider that
$D_r\propto 1/\eta$, and that $\eta\propto (\rho-1)$~\cite{ihl03b} we
can explain the major $D_r$ by considering the rescaled coefficient
$\tilde{D}_r = D_r(\rho|_{L_Z/2}-1)/(\rho-1)$.  The factor
$\rho|_{L_Z/2}$ can be measured in the simulations, or calculated from
the ideal gas equation of state as $\rho/\rho|_{L_Z/2} =
\overline{T}\ln(T_h/T_c)/(T_h-T_c)$~\cite{luese12jcp}. The rescaled
coefficient $\tilde{D}_r$ in Fig.~\ref{fig:Dr} shows to be mostly
independent of the applied temperature gradient within the precision
of the data, which demonstrates that the temperature gradient has no
effect in the particle orientation.

It is interesting to mention that {\em thermomolecular orientation}
has been previously reported in diatomic
fluids~\cite{romer12,daub14,lee16}. 
In these cases, elongated molecules made of two atoms with unequal
sizes show to display certain average orientation towards the
direction of the temperature gradient.  This does not contradict the
discussed lack of orientation induced by anisotropic thermophoresis,
since the particules we discuss are elongated but intrinsically
symmetric, namely composed of indistinguisable building blocks.

\begin{figure}[h!]
\includegraphics[width=0.3\textwidth]{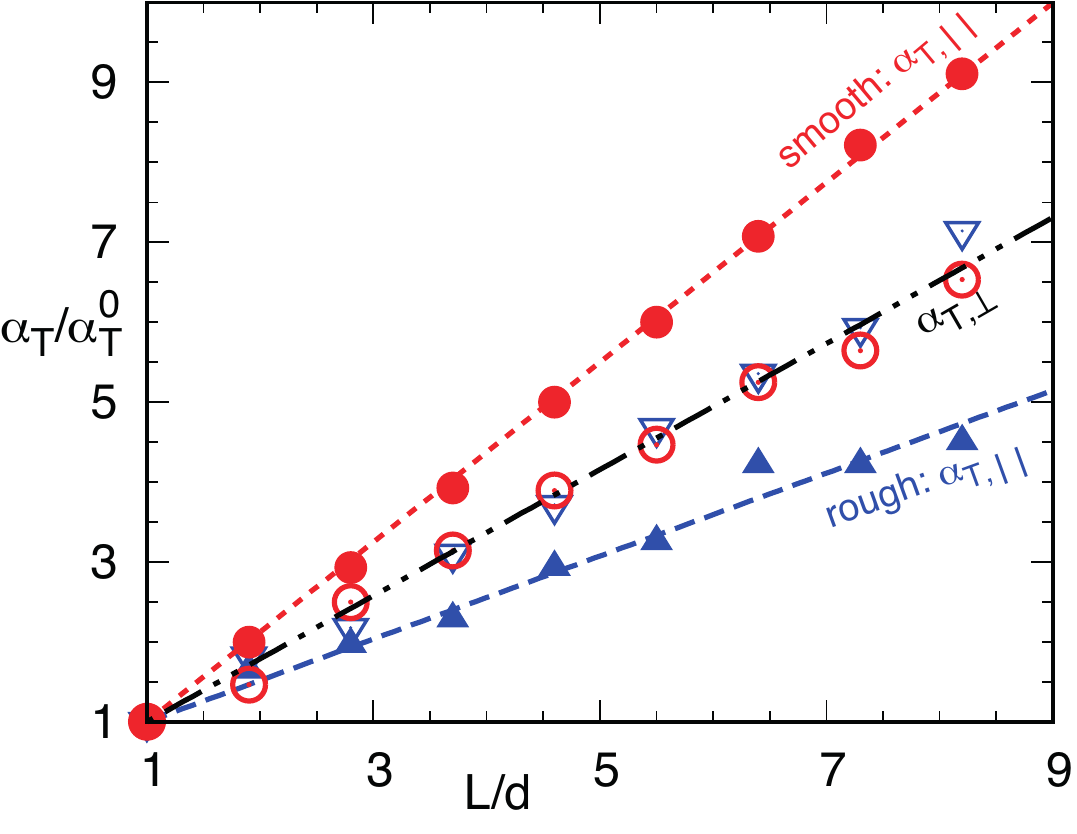}
\caption{\label{fig:aT_aspR} Normalized parallel (open symbols) and
  perpendicular (filled symbols) thermodiffusion factors as a function
  of the aspect ratio $L/d$ for smooth and rough rods (rugosity
  parameters $l/d= 0.3$ and $0.9$) with {\em r6} interaction
  potential.}
\end{figure}

\paragraph{Aspect ratio effect.-}
For spherical colloids, the size is well-known to influence
$\alpha_T$, the particle thermodiffusion
factor~\cite{piazza08,braun06c,luese12jpcm}.  The overall increase of
size of the rod will therefore also translate into an increase of
$\alpha_T$, but how does the rod aspect ratio influence the
anisotropic effect, still needs to be clarified.  Simulations of rods
with orientations parallel and perpendicular to the temperature
gradient are performed for different aspect ratios and for two
rugosities as shown in Fig.~\ref{fig:aT_aspR}.  
Obtained measurements are normalized with $\alpha_{T}^0$, the thermal
diffusion factor of a single bead of the same characteristics as the
ones employed to build the rod, such that for $L=d$,
$\alpha_{T,\perp}=\alpha_{T,\|}=\alpha_{T}^0$ by definition.  The four
analyzed cases in Fig.~\ref{fig:aT_aspR} show a very clear linear
increase of $\alpha_T$ with the aspect ratio $L/d$, although with
different slopes. The linear increase can be easily understood since
the increase of the rod surface exposed to the temperature gradient
also increases linearly. This means that by characterizing the values
of $\alpha_T$ at two different aspect ratios with enough precision,
would allow us to easily extrapolate to other aspect ratios; also in
the case that one of those values is the limiting spherical case
$\alpha_T^0$. 
In the case of a freely rotating colloidal rod, the overall
thermodiffusion factor will therefore also increase linearly with the
aspect ratio of the colloidal rod. This is very interesting by
  itself, and reminiscent of the well-known effect of the particle
  moment of inertia on the thermal diffusion in molecular
  mixtures~\cite{plathe00,koeh01,gal03}.

\begin{figure}[h!]
\includegraphics[width=0.25\textwidth]{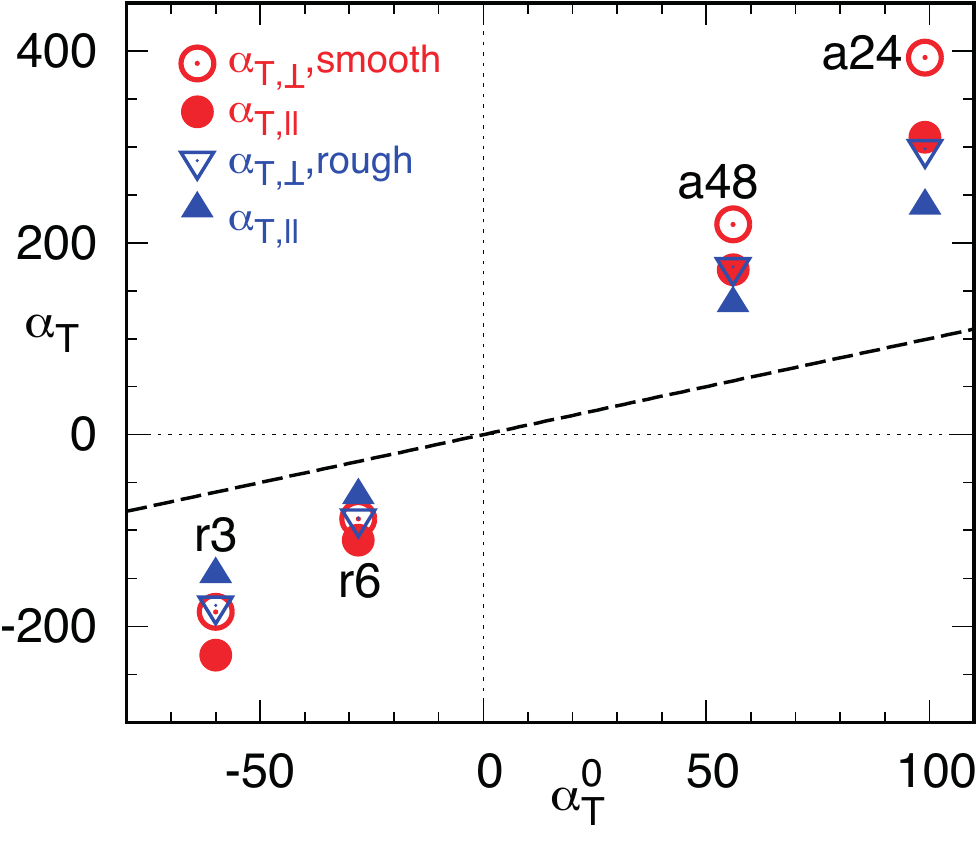}
\caption{\label{fig:app_a0} Parallel and perpendicular thermodiffusion
  factors for rods with $L/d=3.7$ with various interaction potentials
  (indicated in the labels) as a function of $\alpha_T^0$, the
  thermodiffusion factor of a single bead. Results are provided for
  the same two rugosity parameters as in Fig.~\ref{fig:aT_aspR}. The
  dashed line is a guide to the eye.}
\end{figure}

\begin{figure}[h!]
\includegraphics[width=0.49\textwidth]{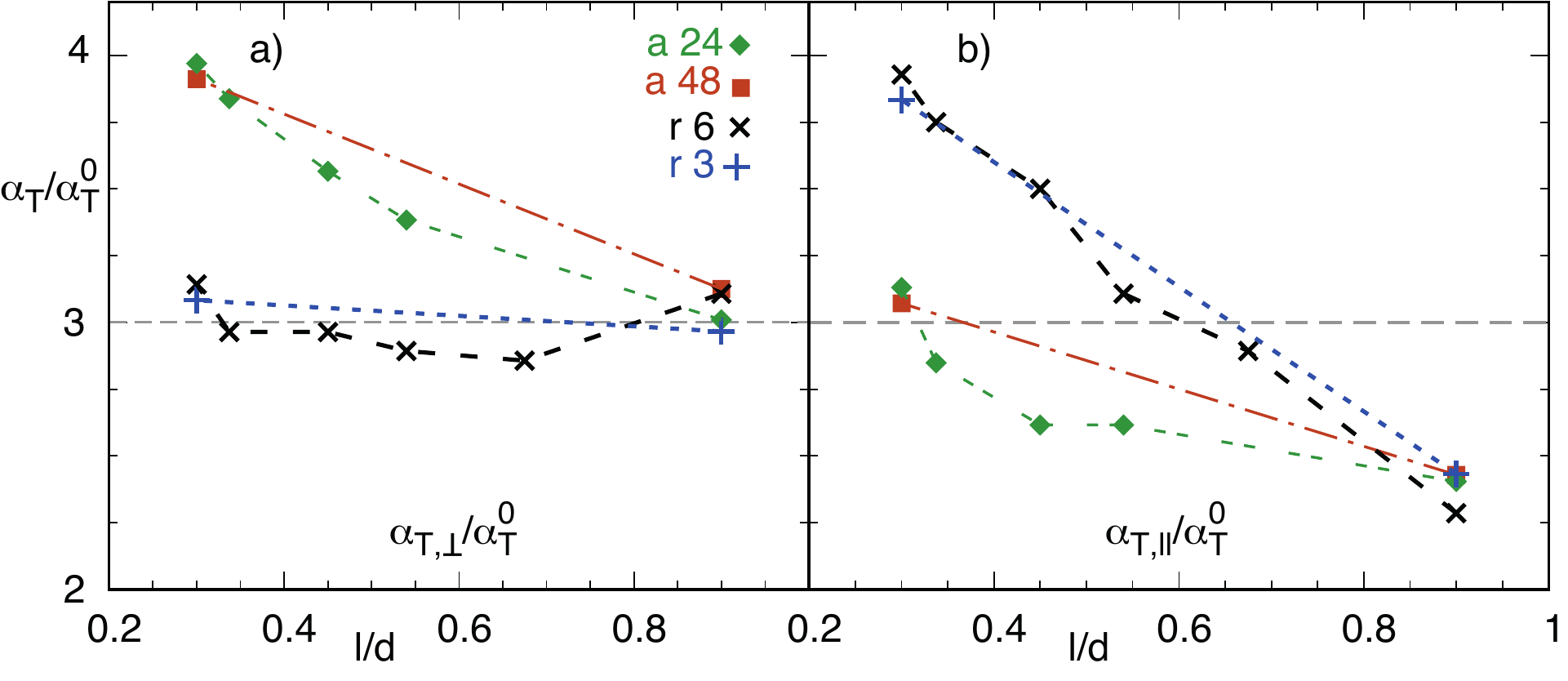}
\caption{\label{fig:aT_pp} Normalized parallel and perpendicular
  thermodiffusion factors for rods with $L/d=3.7$ as a
  function of the rugosity parameter $l/d$.  Simulations with various
  interaction potentials are reported and indicated in the
  corresponding labels. The horizontal line is a guide to the
  eye.}
\end{figure}

\paragraph{Surface  effects.-}
The rod surface is modified in our model in two different manners. On
the one hand, the surface shape as described in Fig.~\ref{model}
modifies its rugosity with the parameter $l/d$. On the other
hand, the choice of the employed potential, attractive-repulsive,
soft-steep will also modify the thermophoretic properties of the
rod. 
To analyze more in detail these effects, further simulations calculate
$\alpha_{T,\|}$ and $\alpha_{T,\perp}$ as a function of the different
potential interactions and the rugosity parameter $l/d$, as shown in
Fig.~\ref{fig:app_a0} and Fig.~\ref{fig:aT_pp}. 
In the case of spherical colloids, the thermodiffusion factor
$\alpha_T^0$ is known to depend on the colloid-surface interactions,
and on the colloid size~\cite{piazza08r,luese12jpcm}. With rods of
fixed aspect ratio, we perform simulations with repulsive and
attractive potentials, and with different $n$ values in
Eq.~(\ref{lj}), for which $\alpha_T^0$ is also evaluated. The parallel
$\alpha_{T,\|}$, and perpendicular $\alpha_{T,\perp}$, thermodiffusion
factors are compared with $\alpha_T^0$, as displayed in
Fig.~\ref{fig:app_a0}. 
Values larger than the reference linear increase in
Fig.~\ref{fig:app_a0}, in absolute numbers, indicate the enhancement
of the thermal diffusion factors due to the increase elongation of the
rod. This enhancement is clear in all investigated cases, although its
magnitude differs for the different potentials, rugosities, and
orientations. In general, the increase of $\alpha_{T,\|}$ and
$\alpha_{T,\perp}$ shows to be larger, the larger the $\alpha_T^0$. 

The effect of the surface rugosity can already be observed in
Fig.~\ref{fig:FT_angle} and Fig.~\ref{fig:aT_aspR} where simulation
results of two well-differentiated rod rugosities are presented.  For
the repulsive potential and aspect ratio here employed, the rough rod
shows $|\alpha_{T,\|}| < |\alpha_{T,\perp}|$; {\em i. e.} the
thermophoretic force for the rod aligned with the temperature gradient
is smaller than for the perpendicular one. Meanwhile the smooth rod
with the same potential shows the opposite behavior, $|\alpha_{T,\|}|
> |\alpha_{T,\perp}|$. This means that by changing the rugosity of the
rod, the forces perpendicular to the temperature gradient can invert
their direction, as shown in Fig.~\ref{fig:FT_angle}. 
This is though not the case for the rods simulated with attractive
interactions where $\alpha_{T,\|} < \alpha_{T,\perp}$ for both the
smooth and the rough surfaces. %
Note that the sign of the thermodiffusive factors is never modified,
such that the thermophilic/thermophobic character of the colloids does
not change with its shape change from spherical into elongated, and it
will be therfore the same for both components $\alpha_{T,\|}$ and
$\alpha_{T,\perp}$. 

The dependence of the thermal diffusion factors with the rugosity
parameter $l/d$, is shown in Fig.~\ref{fig:aT_pp}.  With increasing
rugosity of the rod surface, the thermal diffusion factor shows to
decrease, or to remain unchanged. In other words, smoother surfaces
produce larger thermophoretic responses or eventually similar than
their rough counterparts.  The clear decrease of $\alpha_{T,\|}$ with
increasing rugosity is what eventually changes the relative importance
of both factors and the direction of the force perpendicular to the
temperature gradient.

\begin{figure}[h!]
\includegraphics[width=0.2\textwidth]{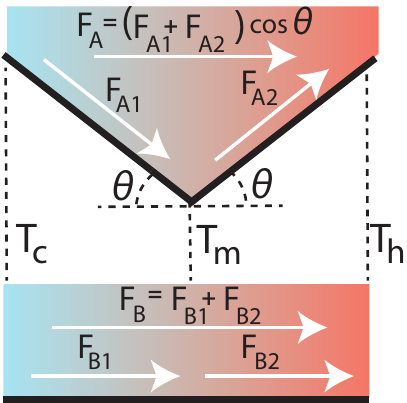}
\caption{Sketch illustrating the decrease of the thermophoretic 
     force along an indented surface.
  \label{fig:indent} }
\end{figure}

To explain these dependencies, we should discuss two types of
contributions, which are again related to the rod-solvent interaction
potentials and the surface shape. The finite range of the solvent-bead
interactions results in overlapping regions, in which solvent
particles can simultaneously interact with more than one neighboring
bead. The size of these overlapping areas decreases with increasing
roughness, or potential steepness, modifying the effective rod-solvent
potential. This effect is smaller for rough rods, for which the
thermodiffusion factors in Fig.~\ref{fig:aT_pp} have little dependence
of the type of potential. Smooth rods interestingly show a difference
between attractive (thermophobic) potentials, and repulsive
(thermophilic) ones, but not between those with different
steepness. The second type of contribution is in this case dominant,
and related to the surface shape, and in particular with the presence
of surface indentations which can explain the decrease of
$\alpha_{T,\|}$ with increasing roughness. To estimate the
contribution to thermophoretic force between two arbitrary points at
the rod surface, we consider first that the thermophoretic factor
along a wall, can be assumed to be directly proportional to the wall
length, which has been shown in Fig.~\ref{fig:aT_aspR}.  By
considering Eq.~(\ref{eq4}), and that the gradient depends on the
inverse of the distance, the contribution to thermophoretic force can
then be determined by the difference of temperatures between these two
points $T_c$ and $T_h$.  Figure~\ref{fig:indent} illustrates the
thermophoretic forces along a wall with an indentation, $F_A$, and a
flat wall $F_B$, which are both considered to be the sum of two
contributions by considering the temperature at the middle of the wall
$T_m$. Indented and flat walls have different lengths, but the
temperatures at their ends are the same. 
The wall length increase exactly cancels with the decrease of the
temperature gradient, such that the wall thermophoretic forces are
precesely the same, this is $F_{A1}=F_{B1}$ and $F_{A2}=F_{B2}$.
The total force in both cases is though not the same due to the angle
$\theta$ that determines the indentation, as sketched in
Fig.~\ref{fig:indent}, such that the force in the indented surface
$F_A$ is a factor $cos\theta$ smaller than the force in the perfectly
smooth surface $F_B= F_A/cos\theta$.
This effect is very clear when the rod is aligned with the temperature
gradient, and explains the clear decrease of $\alpha_{T,\|}$ with
$l/d$ for all potentials in Fig.~\ref{fig:aT_pp}b. When the rod
  is perpendicular to the gradient, the effect of the indentation is
  subtly different but still present due to employed three-dimensional
  structure. However, the overall contribution is smaller which explains
  the unchanged or decreasing dependence of $\alpha_{T,\perp}$ with
  $l/d$ in Fig.~\ref{fig:aT_pp}a.

\paragraph{Thermophoretic anisotropy factor.-}
The importance of the anisotropic effect is determined by how
different are the thermophoretic forces of the rods aligned and
perpendicular to the temperature gradient. We therefore define the
dimensionless {\em thermophoretic anisotropy factor} as
\begin{equation}
  \chi_T = \alpha_{T,\perp}- \alpha_{T,\|}.
\label{chi}
\end{equation}
The intensity and the sign of the force perpendicular to the
temperature gradient is simply determined by $\chi_T$, as
already shown in Eq.~(\ref{perp}). The direction of the perpendicular
force will have crucial importance in applications of the effect,
determining for example the rotation direction of the thermophoretic
turbines~\cite{yang14turb}.  
This means that the sign of $\alpha_T^0$ will not be enough to know
the direction of the perpendicular force; or, in other words, the
thermophilic or thermophobic character of the surface does not
determine the direction of the transverse phoretic effect. 

\begin{figure}[h!]
\includegraphics[width=0.5\textwidth]{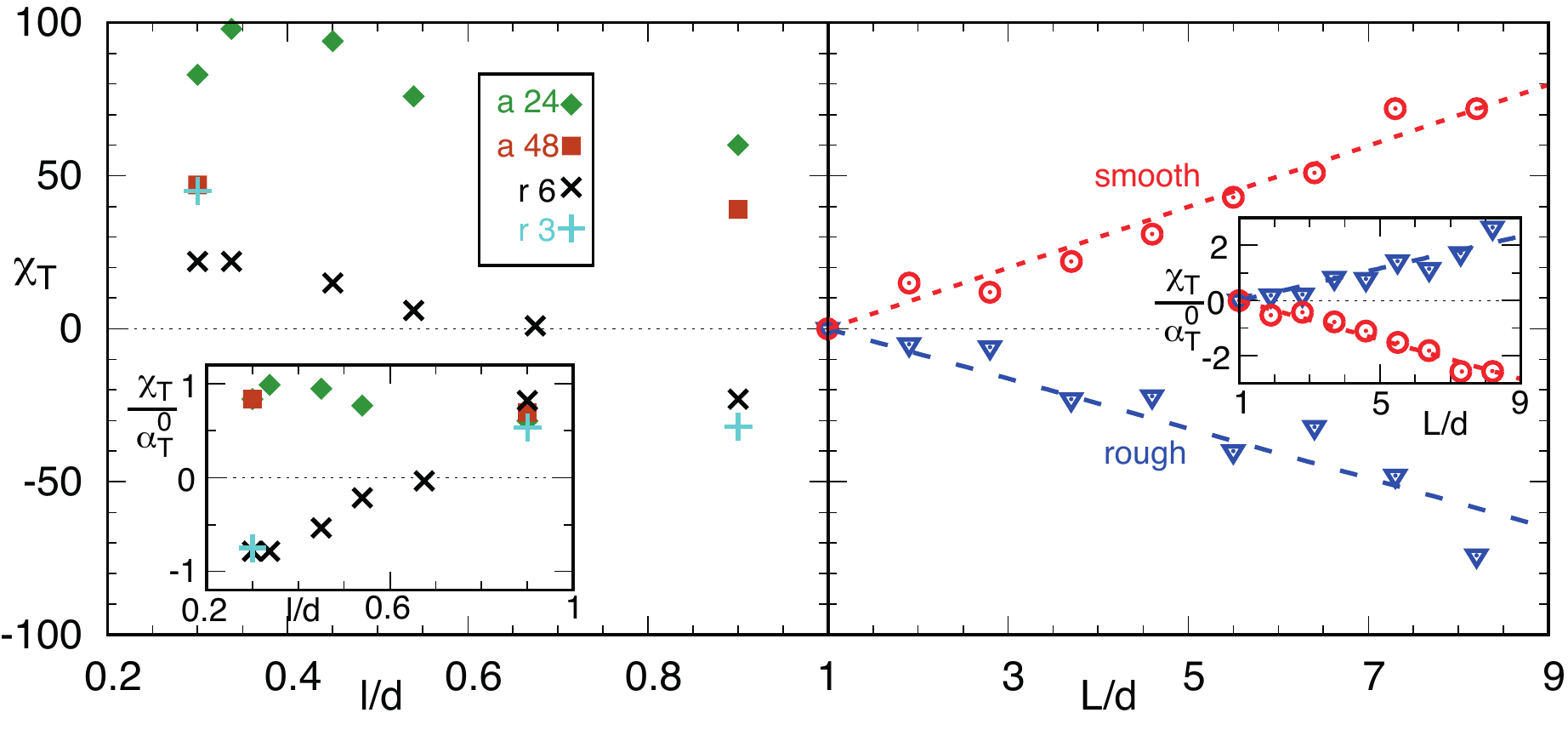}
\caption{a)~Asymmetric thermophoretic factor $\chi_T$ as
  function of the rod rugosity parameter, for various interaction
  potentials, with $L/d=3.7$. 
  b)~$\chi_T$ as a function of the aspect ratio for {\em r6} interaction
  potential and rugosities  $L/d=0.3$ and 0.9. 
  The insets correspond to the ratio $\chi_T/\alpha_T^0$ for the same data. 
  \label{fig:FT_chi} }
\end{figure}
The value of $\chi_T$ is displayed for various potentials and
rugosities in Fig.~\ref{fig:FT_chi}, where it can be observed that
$\chi_T>0$ in the majority of the cases. This means that the
perpendicular force happens most of the time in the same direction,
which is positive with the convection in
Fig.~\ref{fig:t-anisotrop}. 
In the thermophobic case, where the thermodiffusion factor is
positive, $\chi_T>0$ corresponds to $|\alpha_{T,\perp}| >
|\alpha_{T,\|}|$ what resembles the well-know translational friction
case.  However, in the thermophilic case, where the thermodiffusion
factors are negative $\chi_T>0$ corresponds to $|\alpha_{T,\perp}| <
|\alpha_{T,\|}|$. For comparison, the insets in Fig.~\ref{fig:FT_chi}
show the ratio $\chi_T/\alpha_T^0$, which will be positive in the
friction-like cases, and negative otherwise. 
This normalization also provides a rapid estimation of the magnitude
of the effect which can be as expected to be as large as the phoretic
effect itself, even for the small aspect ratios here considered. 
Note though that these are the results obtained by means of computer
simulation and that in practical experiments, the results can show a
much richer behavior.  With our simulation results, we observe that by
changing the surface rugosity, the anisotropic effect can reverse its
direction in the case of thermophilic rods. We expect this behavior to
be reproducible experimentally by changing the surface coating,
electrostatic interactions, average temperature, or any of the factors
that are known to affect the thermophoretic behavior.

\section{Conclusions}
Anisotropic thermophoresis refers to the different phoretic thrust
that an elongated particle suffers when aligned with the temperature
gradient and when perpendicular to it.  This difference results in a
contribution to the thermophoretic force perpendicular to the
temperature gradient when the rod is fixed oblique to the gradient. 
This anisotropy does not have any relevant effect on the particle
orientation, nor on the rotational diffusion of the particle, given
the considered symmetry of the colloidal rod. 
Although unreported until know, the existence of anisotropic
thermophoresis is relatively intuitive, especially by comparison with
the translational friction of a rod which is also noticeable different
if aligned with the flow, or perpendicular to it.  
Here we analyze this effect in detail, showing that the intensity and
the direction of the force is a function of the aspect ratio, the
surface geometry, and the colloid-fluid interactions. Increasing
aspect ratio linearly increases the anisotropic phoretic effect in an
straightforward manner. The rugosity of the colloidal surface is also
relevant, being smooth surfaces the ones with larger anisotropic
thermophoretic effect. In general, surfaces with larger phoretic
effect also have larger anisotropic phoretic effect. Interestingly,
and for the simulation potentials employed in this work, the direction
of the phoretic force perpendicular to the gradient is the same for
both thermophobic and thermophilic colloids.  Only rods with
thermophilic and rough surface have shown in our simulations to
display the perpendicular forces with opposite direction. We expect
that experimental results will show an even richer behavior, in which
the intensity and the direction of the effect could depend on many
factors such as intrinsic surface properties, eventual coatings,
average fluid temperature and density, presence of salt ions, and
various other factors. 
Two first practical applications of anisotropic thermophoresis have
already been found in the construction of phoretic microturbines and
micropumps. In the presence of external temperature gradients, the
blades of a microturbine will rotate when being
anisotropic~\cite{yang14turb}, and a microchannel will experience some
spontaneous directed fluid motion when including elongated tilted
obstacles~\cite{tan2}. 
Although the work presented here has been exclusively focused in the
thermophoretic effect, very similar results are expected with other
phoretic effect such as diffusiophoresis. A direct proof for this is
the fact that a similar micro-turbine placed in an external
concentration gradient has shown to display similar behavior due to
the related anisotropic diffusiophoresis~\cite{yang15cturb}. 
In summary, with this investigation, we provide a deep insight into
the anisotropic thermophoresis of elongated micro-meter size objects;
effect that we hope will be soon experimentally verified, and find
applications in different fields like particle characterization,
microfluidics, or biomedicine.

\section*{Acknowledgments}
We gratefully thank the computing time granted on the supercomputer
JURECA at J{\"u}lich Supercomputing Centre (JSC). Financial support from China Scholarship Council
(CSC) is gratefully acknowledged. M.~Y. also acknowledges financial support of the NSFC
(No. 11674365).

\end{document}